\documentclass[useAMS,usenatbib,usegraphicx,letters]{mn2e}
\usepackage{natbib,epsfig,graphicx,rotating,lscape,rotating,amssymb,color,subfigure,times}
\usepackage[usenames,dvipsnames,svgnames,table]{xcolor}

\def\h2o{\rm H_{2}O}
\def\ch4{\rm CH_{4}}
\def\co2{\rm CO_{2}}

\title[Day side water absorption in HD 189733 b]{Detection of water
  absorption in the day side atmosphere of HD 189733 b using
  ground-based high-resolution spectroscopy at $3.2\mu$m\thanks{Based
    on observations collected at the European Southern Observatory
    (186.C-0289)}} \author[J. L. Birkby et
al.]{J. L. Birkby$^{1}$\thanks{E-mail: birkby@strw.leidenuniv.nl
    (JLB)}, R. J. de Kok$^{2}$, M. Brogi$^{1}$,
  E. J. W. de Mooij$^{3}$, H. Schwarz$^{1}$, \newauthor S. Albrecht$^{4}$, I. A. G. Snellen$^{1}$\\\\
  $^{1}$Leiden Observatory, Leiden University, Niels Bohrweg 2, 2333
  CA Leiden, The Netherlands\\
  $^{2}$SRON Netherlands Institute for Space Research, Sorbonnelaan 2,
  3584 CA Utrecht, The Netherlands\\
  $^{3}$Department of Astronomy and Astrophysics, University of
  Toronto, 50 St. George Street, Toronto, ON M5S 3H4m, Canada\\
  $^{4}$Department of Physics, and Kavli Institute for Astrophysics
  and Space Research, MA Institute of Technology,
  Cambridge, Massachusetts 02139, USA}

\begin{document}
\date{Accepted 2013 August 1. Received 2013 July 23; in
  original form 2013 July 4}
\pagerange{\pageref{firstpage}--\pageref{lastpage}}
\maketitle
\label{firstpage}

\begin{abstract}
  We report a $4.8\sigma$ detection of water absorption features in
  the day side spectrum of the hot Jupiter HD 189733 b. We used
  high-resolution ($R\sim100~000$) spectra taken at $3.2~\mu$m with
  CRIRES on the VLT to trace the radial-velocity shift of the water
  features in the planet's day side atmosphere during 5 h of its 2.2 d
  orbit as it approached secondary eclipse. Despite considerable
  telluric contamination in this wavelength regime, we detect the
  signal within our uncertainties at the expected combination of
  systemic velocity ($V_{\rm sys}=-3^{+5}_{-6}$ km s$^{-1}$) and
  planet orbital velocity ($K_{\rm p}=154^{+14}_{-10}$ km s$^{-1}$),
  and determine a $\h2o$ line contrast ratio of
  $(1.3\pm0.2)\times10^{-3}$ with respect to the stellar continuum. We
  find no evidence of significant absorption or emission from other
  carbon-bearing molecules, such as methane, although we do note a
  marginal increase in the significance of our detection to
  $5.1\sigma$ with the inclusion of carbon dioxide in our template
  spectrum. This result demonstrates that ground-based,
  high-resolution spectroscopy is suited to finding not just simple
  molecules like CO, but also to more complex molecules like $\h2o$
  even in highly telluric contaminated regions of the Earth's
  transmission spectrum. It is a powerful tool that can be used for
  conducting an immediate census of the carbon- and oxygen-bearing
  molecules in the atmospheres of giant planets, and will potentially
  allow the formation and migration history of these planets to be
  constrained by the measurement of their atmospheric C/O ratios.
\end{abstract}

\section{Introduction}

In the past three years, high-resolution, near-infrared, ground-based
spectroscopy has identified the signature of molecular absorption by
carbon monoxide (CO) in the atmospheres of several hot Jupiters,
including in the transmission spectrum of HD 209458 b \citep{Sne10},
and in the thermal day side spectra of the transiting planet HD 189733
b \citep{deK13,Rod13}, and the non-transiting planets $\tau$ Bo\"otis
b \citep{Brog12,Rod12} and tentatively 51 Pegasi b \citep{Brog13}. The
more significant of these detections have been made with the CRyogenic
high-resolution InfraRed Echelle Spectrograph (CRIRES;
\citealt{Kae04}) on the Very Large Telescope (VLT) at a resolution of
$R\sim100~000$ targeting the individual lines of the CO band head at
$2.3~\mu$m. The large change in the radial velocity of the planets
($\sim100$ km s$^{-1}$) during their orbits allows their spectra to be
disentangled from the essentially stationary lines of their host stars
and from the Earth's static telluric lines. A simple cross-correlation
of the extracted planet spectrum with models of CO transitions for
different atmospheric temperature--pressure ($T/P$) profiles and
volume mixing ratios (VMRs) not only revealed the presence of CO in
the planetary atmosphere, but also allowed the planet's orbital
velocity and hence its orbital inclination to be calculated. In the
cases of the transiting planets, this allowed them to be treated as
eclipsing binary systems, resulting in model-independent measurements
of the true masses and radii of the host star and
planet. Ground-based, high-resolution spectroscopy is clearly a
powerful technique for characterizing exoplanets and their atmospheres
\citep{Sne13}, but its potential for detecting other molecules, in
particular the other main carbon- and oxygen-bearing species, such as
water ($\h2o$), methane ($\ch4$) and carbon dioxide ($\co2$), is as
yet untested, particularly in more opaque regions of the Earth's
atmosphere. Ultimately, the technique can be used to provide
constraints on the relative abundances of these molecules in planetary
atmospheres, and hence an estimate of the carbon-to-oxygen ratio
(C/O), which is thought to have strong implications for the formation
and migration history of the planets (see e.g. \citealt{Lod04,Obe11}).

HD 189733 b is one of the most studied exoplanets to date, with strong
evidence for a high-altitude haze that causes Rayleigh scattering from
$0.3$ to $1~\mu$m \citep{Pon08,Sin09,Sin11,Pon13}, reports of $\h2o$
absorption features at infrared wavelengths \citep{Tine07,Gri08,Swa08}
and claims of $\ch4$ fluorescence at $3.25~\mu$m
\citep{Swa10,Wal12}. However, there is some interesting debate in the
literature about the latter two spectral features with both
systematics and the possible haze being proposed as the causes of
conflicting results at different wavelengths
\citep{Gri07,Ehr07,Des09,Sin09,Gib11,Mandell11,Gib12}. In this Letter,
we present $R\sim100~000$ time-resolved CRIRES spectra of the hot
Jupiter HD 189733 b, centred on $3.2~\mu$m, targeted at detecting the
potential molecular signatures of $\ch4$, $\h2o$ and also $\co2$ in
the planetary atmosphere.  Our choice to observe at $3.2~\mu$m was
driven by the claimed detection of methane fluorescence in this region
for HD 189733 b, which would produce easily identifiable emission
features in the residuals of our high-resolution spectra given the
$\sim1$ per cent emission features seen at much lower resolution by
\citet{Swa10}. However, the $3.2~\mu$m region probed by CRIRES suffers
almost total telluric absorption in some parts (unlike previous
observations at $2.3~\mu$m) which has the potential to degrade the
results of the cross-correlation technique as there will be fewer
pixels to use in the analysis. In addition, the molecular spectra of
$\h2o$, $\ch4$ and $\co2$ are far more complex than the CO spectra
used in our previous analysis, with many lines that are extremely weak
at the temperatures accessible to laboratory measurements. Accurate
line positioning in the models is key to the success of the
cross-correlation technique, but ab initio calculations are necessary
to generate the hot model spectra we require. This may result in small
errors in the line positions \citep{Bai12}, but water vapour lines are
well constrained by observations \citep{Barber06}.

\section{Observations and data reduction}
\subsection{Observations}
We observed HD 189733 (K$1$V, $V=7.68$ mag, $K=5.54$ mag) as part of
the large ESO programme 186.C-0289, which was designed to detect the
spectral signatures of molecular species in the atmospheres of the
brightest known transiting and non-transiting systems accessible from
Chile. We observed the target for $\sim5$ h during the night of 2011
August 1, using CRIRES mounted at Nasmyth A focus on the 8.2-m
telescope UT1 (Antu) of the VLT, located on Cerro Paranal in
Chile. The observations were carried out in combination with the
Multi-Application Curvature Adaptive Optic system (MACAO;
\citealt{Ars03}) and a $0.2$ arcsec slit centred on $3236$ nm (order
17). CRIRES consists of four Aladdin III InSb-arrays each spanning
$1024\times512$ pixel, with a gap of $\sim280$ pixel between each
chip. The resulting wavelength coverage of our observations was thus
$3.1805<\lambda(\mu$m$)<3.2659$ with a resolution of $R\sim100~000$
per resolution element. The planet was observed without interruption
between orbital phases of $0.383<\phi<0.475$ as the maximum day side
illumination of the planet was rotating into view, corresponding to a
total planet radial-velocity change of $\sim75$ km s$^{-1}$. In total,
we obtained 48 spectra, with each spectrum consisting of two sets of
$5\times30$ second exposures.  To allow for accurate sky-background
subtraction, the telescope was nodded along the slit by $10$ arcsec
between each set of exposures in an ABBA sequence. A standard set of
calibration frames was taken the following morning.

\subsection{Basic data reduction}
\begin{figure*}
\centering
\includegraphics[width=0.975\textwidth]{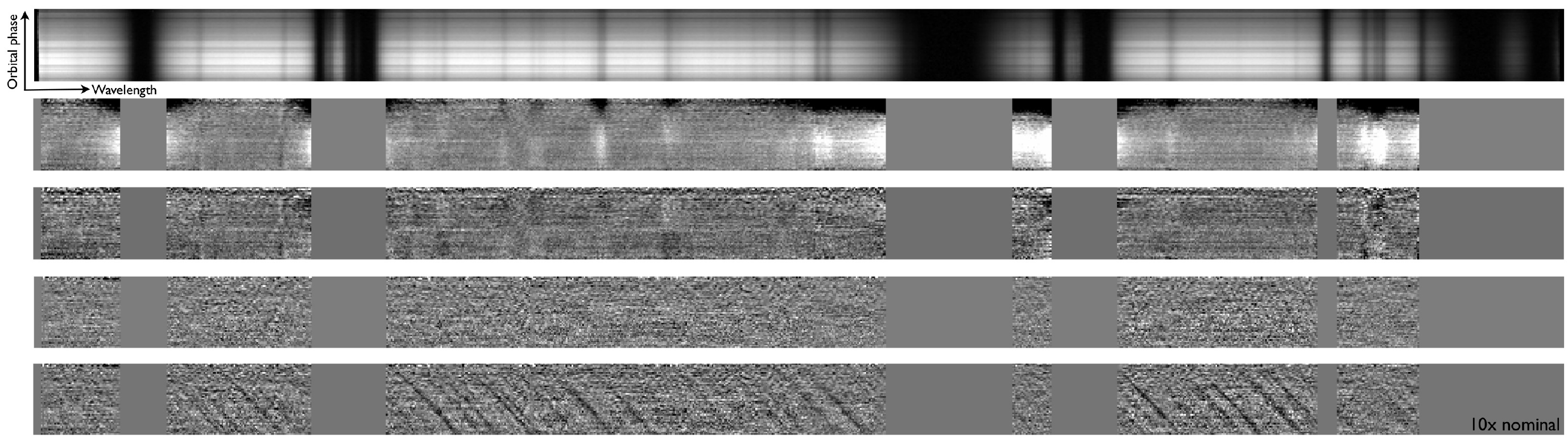}
\caption{Removal of telluric features on detector 1 by {\sc Sysrem}
  iterations. The top panel shows the CRIRES pipeline spectra after
  aligning to a common wavelength grid. Time (or frame number)
  increases vertically on the $y$-axis of the matrices, while pixel
  number increases along the $x$-axis. The second panel shows the
  residuals of the mean-subtracted spectra given as the first input to
  {\sc Sysrem}.  The greyed-out regions show where we applied the mask
  for near total telluric absorption. The third panel shows the
  residuals after the first iteration of {\sc Sysrem} which has
  removed a trend that correlates strongly with airmass. The fourth
  panel shows the residuals after the optimal number of {\sc Sysrem}
  iterations for detector 1 (8 in this case) and after division of
  each pixel by the squared standard deviation of its column. The
  standard deviation of this matrix is $4.5\times10^{-3}$. For
  reference, the bottom panel shows the same as the fourth panel but
  with the best-matching cross-correlation template injected at $10$
  times the nominal value before running {\sc Sysrem}, to highlight
  how the planetary lines shift during the night compared to the
  telluric features.}
\label{fig:sysrem}
\end{figure*}

We carried out the initial two-dimensional (2D) image processing and
extraction of the 1D spectra using version 2.2.1 of the CRIRES {\sc
  esorex} pipeline. The data were flat-fielded and corrected for bad
pixels and non-linearity effects, then background-subtracted by
combining each AB nodding pair, before using an optimal extraction
technique \citep{Hor86} to obtain the 1D spectra. The pipeline
products require post-processing in order to remove the contaminating
telluric features. For this purpose, we used a combination of {\sc
  iraf} routines and custom-built {\sc idl} procedures. Each CRIRES
detector is read out using a different amplifier, and each has its own
particular characteristics that need to be dealt with
independently. Consequently, we handled the 1-D spectra from each
detector separately, creating four matrices of size $1024\times N$,
where $N$ is the number of spectra, sorted in order of time
(i.e. phase) along the $y$-axis, while the $x$-axis corresponds to
pixel number (i.e. wavelength). An example of the matrix created for
detector 1 can be seen in the top panel of Fig.~\ref{fig:sysrem}.

Our first post-processing step was to mask any groups of bad columns
in the matrices, i.e. those typically associated with detector defects
at the beginning and end of each detector. We then performed an
additional bad-pixel correction to fix bad regions and pixels not
identified by the pipeline. Singular bad pixels and isolated bad
columns were identified by eye. The bad pixels were replaced with
spline-interpolated values from their horizontal neighbouring
pixels. Additional residual bad pixels were identified iteratively
during this process, with a total of $0.2-0.9$ per cent of the pixels
in each matrix requiring correction.
Next, we selected the spectrum in each matrix with the highest
signal-to-noise ratio (S/N) as a reference and used it to align all of
the spectra on to a common wavelength grid in pixel space. To do this,
we made use of the stationary telluric features in the spectra and
performed a cross-correlation between each spectrum and the reference
using the {\sc iraf} task {\sc fxcor}. The measured pixel offsets from
the reference were applied to each spectrum using a global spline
interpolation to align them with the reference spectrum.

We derived a common wavelength solution by identifying the wavelengths
of the telluric features in the reference spectrum based on comparison
with a synthetic telluric transmission spectrum from
ATRAN\footnote{http://atran.sofia.usra.edu/cgi-bin/atran/atran.cgi}
\citep{Lor92}. The precipitable water vapour (PWV) content that best
represented the atmospheric conditions during our observations was
PWV$=2.0$ mm. The synthetic spectrum was used to create a line list to
pass to the {\sc iraf} function {\sc identify}, which we made by
selecting the minimum data point in each telluric absorption line of
the synthetic spectrum. The {\sc identify} procedure was then used to
mark the pixel positions of the selected telluric features in the
reference spectrum and a wavelength solution in pixel space was
derived using a third-order Chebyshev polynomial. This was used to
update the default pipeline wavelength solution.

\subsection{Removal of telluric contamination with {\sc Sysrem}}
\begin{figure}
\centering
\includegraphics[width=0.4\textwidth]{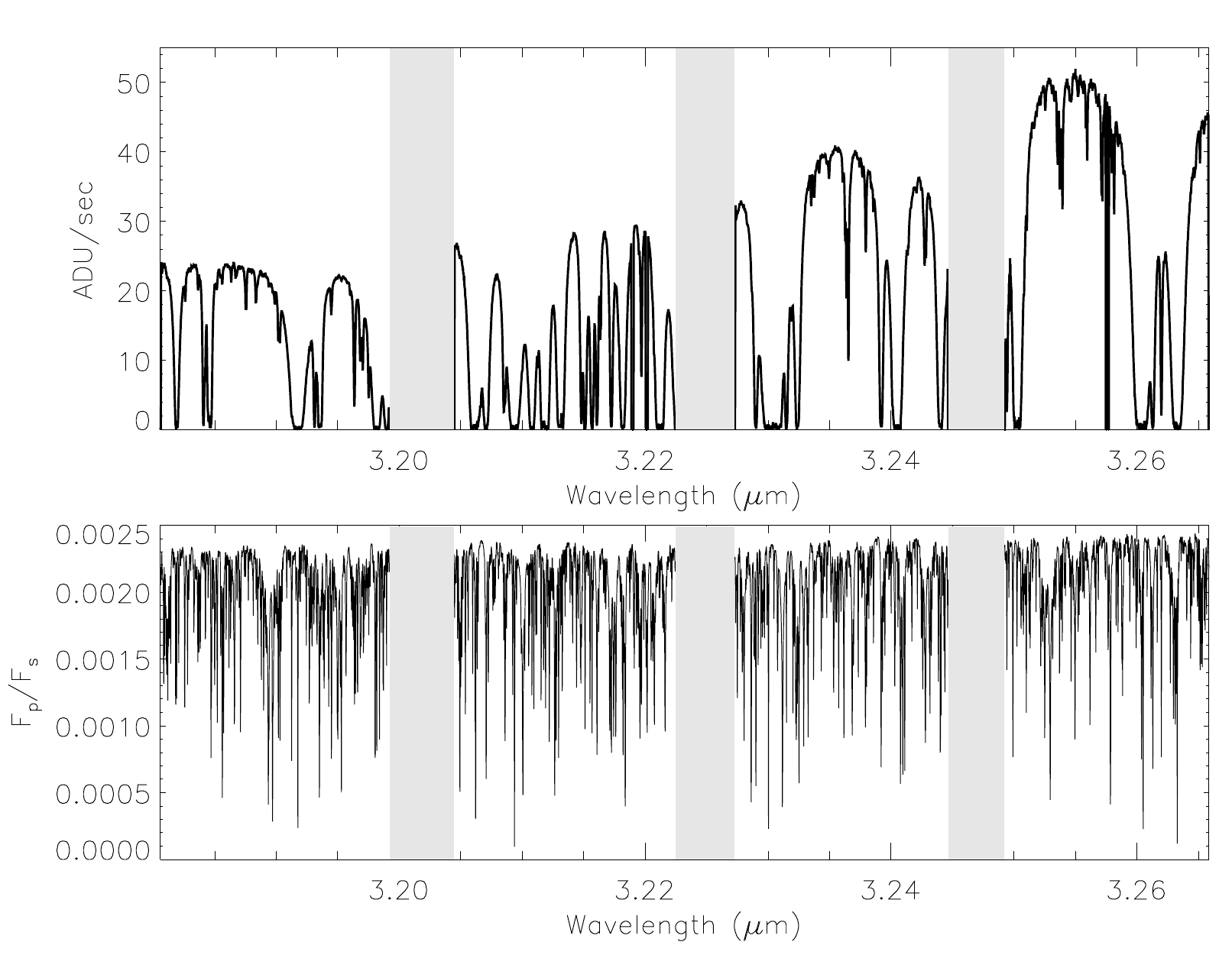}
\caption{{ Top:} an example of a 1D reduced spectrum from the CRIRES
  pipeline before removal of the telluric features. The shaded regions
  mark the gaps between the detectors. Detector 2 suffers significant
  contamination from water in the Earth's atmosphere. { Bottom:} the
  template spectrum of water and carbon dioxide that gave the best
  cross-correlation value. It is noticeably more complex than the
  model spectrum used to detect CO (see fig. 7 of \citealt{deK13}).}
\label{fig:telluric}
\end{figure}

The $3.2~\mu$m region contains many water absorption lines (see
Fig.~\ref{fig:telluric}), and the expected depth of these lines in the
atmosphere of HD 189733 b, with respect to the stellar continuum, is
$\sim10^{-3}$ \citep{Dem06,Char08,Gri08}. Our observed spectra have a
typical S/N of $\sim200$ in the continuum, so the individual water
lines of the planet spectrum are buried in the noise of the data. In
order to extract the planet signal, we used a cross-correlation
technique to combine the contributions from the individual lines
\citep{Brog13,deK13}. However, before we can do this, we must first
remove the dominant signal of telluric contamination (see
Fig.~\ref{fig:sysrem}). The telluric features remain stationary over
the course of the observations and appear as vertical lines in the
matrices. However, they change in strength throughout the night due to
the varying geometric airmass and fluctuations in the water vapour
content of the atmosphere above Paranal. The spectral fingerprint of
the planetary atmosphere on the other hand will be Doppler shifted by
$10$s of km s$^{-1}$ during the night and will trace out diagonal
absorption features across the matrices (see the bottom panel of
Fig.~\ref{fig:sysrem}). In this work, we take a slightly different
approach to removing the telluric contamination than in our previous
studies as part of our ongoing study to optimise the data
reduction. Here, we build upon the method of singular value
decompositions (SVDs) used by \citet{deK13} to identify carbon
monoxide absorption in high-resolution spectra of the day side of HD
189733 b at $2.3~\mu$m.  We have employed the {\sc Sysrem} algorithm
\citep{Tam05,Maz07}, which is commonly used by transit surveys to
de-trend light curves. {\sc Sysrem}, like SVDs, is able to remove
systematic trends without any prior knowledge of the underlying cause,
but has been demonstrated to be more effective in cases where the
errors per data point are not equal \citep{Tam05}. This is
particularly relevant for our $3.2~\mu$m data set due to the broad and
deep telluric absorption lines. In our case, we treat each column (or
wavelength channel) of the spectral matrix as a `light curve'
consisting of $48$ frames. The individual uncertainties on each data
point in the matrix are the error calculated by the optimal extraction
routine of the CRIRES data reduction pipeline for each pixel in each
spectrum. Before executing {\sc Sysrem} on a per detector basis, we
first normalized the spectra to their peak continuum value per
detector and masked regions of almost total telluric
absorption. Finally, we divided each individual spectrum by its mean
pixel value and subtracted unity. An example of the input matrix to
{\sc Sysrem} is shown in Fig.~\ref{fig:sysrem}. The first systematic
component removed by {\sc Sysrem} tightly correlates with air mass for
all four detectors, but subsequent trends do not obviously match with
other physical parameters such as seeing or pressure. In order to
determine the optimal number of {\sc Sysrem} iterations to execute, we
test which combination of iterations and detectors give the highest
significance at the expected planet position. In total, we ran 20
iterations of {\sc Sysrem} per detector. We found that detectors 2 and
4 did not increase the detection significance for any number of the
tested iterations. This is perhaps not surprising for detector 2 given
the heavy masking we applied to the near total telluric absorption
features (see the top panel of Fig.~\ref{fig:telluric}), which left
little signal to work with. Detector 4 is known to suffer reduced
quality due to known variations in the gain between neighbouring
columns (the odd--even effect) caused by the alignment position of the
detector, and such issues have prevented the use of detector 4 in some
of our previous observations \citep{Brog13}.  The effect is a zig-zag
pattern in the 1D spectra on detector 4 which is static in time with
an average amplitude of $\pm5$ per cent around the continuum, but
which scales strongly with increasing count level, peaking at $\sim10$
per cent in some frames. The optimum combination was to use only
detectors 1 and 3, with 8 {\sc Sysrem} iterations on detector 1 and
just one iteration on detector 3. The greater number of iterations
required for detector 1 compared to detector 3 may again be due to the
odd--even effect as it has the same alignment as detector 4. However,
the lower count level on detector 1 compared to detector 4 reduces the
effect to almost negligible levels. As a final step before
cross-correlation, we divide each pixel by the squared standard
deviation of its column.

\section{Cross-correlation analysis and results}

\begin{figure}
\centering
\includegraphics[width=0.44\textwidth]{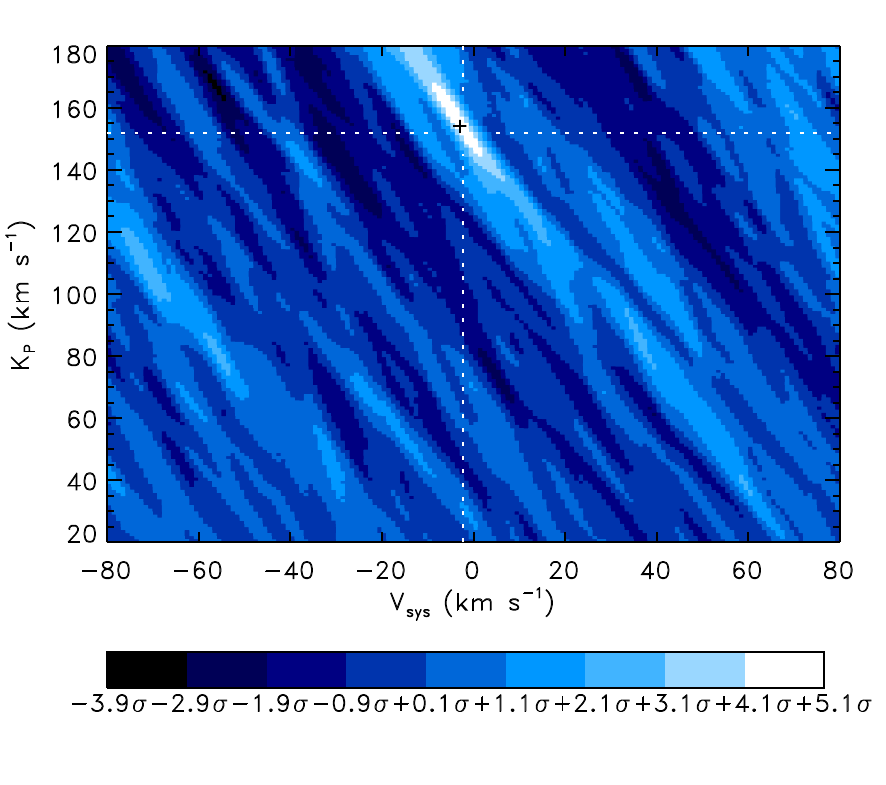}
\caption{Total cross-correlation values from detectors 1 and 3 after
  summing over time for a range of systemic velocities ($V_{\rm sys}$)
  and planet radial velocity semi-amplitude ($K_{\rm p}$). The dashed
  white line marks the expected planet signal based on literature
  values ($V_{\rm sys}=-2.361$ km s$^{-1}$, $K_{\rm p}=152$ km
  s$^{-1}$), while the black plus sign marks the position of the
  maximum cross-correlation value ($V_{\rm sys}=-3^{+5}_{-6}$ km
  s$^{-1}$, $K_{\rm p}=154^{+14}_{-10}$ km s$^{-1}$), which is
  consistent with the literature values within our uncertainties.  The
  white contour marks the $1\sigma$ region around the peak
  cross-correlation value.}
\label{fig:xcor}
\end{figure}

\begin{figure}
\centering
\includegraphics[width=0.37\textwidth]{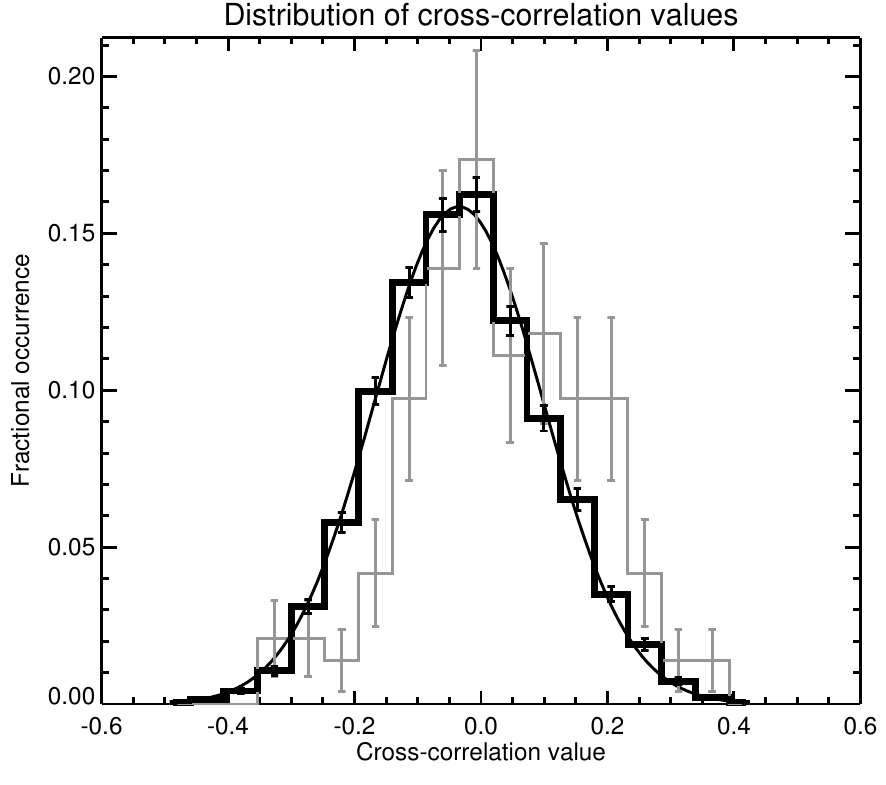}
\caption{A comparison of the in-trail (grey histogram) and
  out-of-trail (black histogram) cross-correlation values. The error
  bars are the square root of the bin occurrences. The out-of-trail
  values are well fitted by a Gaussian (black curve) and the in-trail
  values are offset towards higher cross-correlation values. A Welch
  $T$-test rejects the hypothesis that the two distributions are drawn
  from the same parent population at the $4.9\sigma$ level.}
\label{fig:trail}
\end{figure}

The residuals of each spectrum after running {\sc Sysrem} were
cross-correlated with a grid of models convolved to the CRIRES
spectral resolution containing molecular signatures of different
combinations of $\ch4$, $\h2o$ and $\co2$. The models were generated
for the $3.2~\mu$m region in the same way as those used to study the
$2.3~\mu$m region in \citet{deK13}.
At high pressures ($\geq0.1$ bar), the atmospheric temperature was set
to $1350$K and it then followed the profiles of \citet{Mad09}. For a
lower pressure ($p_{1}$), the temperature ($t_{1}$) was varied from
$500$ to $1500$ K in steps of $500$ K, which allowed for a weak
thermal inversion at high altitudes. Between $0.1$ and $p_{1}$ bar we
assumed a constant rate of change of temperature with
$\log$(pressure), and varied $p_{1}$ between $10^{-1.5}$ and $10^{-4}$
in steps of $10^{0.5}$. The VMRs of the gases were allowed to vary
between $10^{-6}$ and $10^{-3}$ also in steps of $10^{0.5}$. The
cross-correlation analysis was performed over a range of lag values
corresponding to planet radial velocities of $-100\leq$ RV$_{\rm
  p}\leq+200$ km s$^{-1}$. As in our previous studies with CRIRES, the
maximum cross-correlation signal is found by shifting the
cross-correlation functions for each spectrum to the rest frame of the
planet and summing over time for a range of planet radial-velocity
semi-amplitudes ($20\leq K_{\rm p}\leq180$ km s$^{-1}$). Based on
literature values of the planet and host star masses
(e.g. \citealt{Tri09}) and the known inclination of the transiting
system, the expected planet radial velocity is $K_{\rm p}\sim152$ km
s$^{-1}$ at $V_{\rm sys}=-2.361$ km s$^{-1}$ \citep{Bou05b}. The
best-matching cross-correlation template contained both $\h2o$ and
$\co2$ absorption lines, with $t_{1}=500$ K, $p_{1}=10^{-1.5}$,
VMR$_{\rm H_{2}O}=10^{-5}$ and VMR$_{\rm CO_{2}}=10^{-4}$. However,
the detection significance across the full range of temperatures,
pressures and VMRs tested for the $\h2o+\co2$ templates was always
within $1\sigma$ of the best-matching model, which is shown (before
convolution to the CRIRES spectral resolution) in the bottom panel of
Fig.~\ref{fig:telluric}. The strength of the cross-correlation signal
decreased with the inclusion of $\ch4$ in all cases. A matrix
containing the total combined cross-correlation values for the best
$\h2o+\co2$ model is shown in Fig.~\ref{fig:xcor} as a function of
$V_{\rm sys}$ and $K_{\rm p}$. The peak value of the cross-correlation
matrix is located at $V_{\rm sys}=-3^{+5}_{-6}$ km s$^{-1}$ and
$K_{\rm p}=154^{+14}_{-10}$ km s$^{-1}$, which is consistent with
literature values for the expected planet position
\citep{Bou05b,deK13}. We determine the significance of the detection
by dividing the peak value of the cross-correlation matrix by the
standard deviation of the whole matrix, which results in a detection
significance of $5.1\sigma$ for the combined signal of detectors 1 and
3 (individually the two detectors give $4.5\sigma$ and $3.0\sigma$,
respectively). This approach assumes that the distribution of the
cross-correlation values is Gaussian, which is reasonable, despite
possible systematics in the observed spectra, because we have (i)
normalized each pixel by its uncertainty and (ii) by cross-correlating
with a template of many lines that span the entire wavelength range,
systematic variations from a Gaussian distribution are heavily
down-weighted. However, to test the assumption of Gaussianity, we show
the distributions of the cross-correlation values inside and outside
the planet radial-velocity trail in Fig.~\ref{fig:trail}. The
out-of-trail cross-correlation values are well fitted by the Gaussian
curve shown in the plot, and the in-trail values are notably offset
from the out-of-trail distribution.  A Welch $T$-test rules out the
in-trail and out-of-trail values having been drawn from the same
parent distribution at the $4.9\sigma$ level.

\section{Discussion}

The inclusion of $\co2$ in the cross-correlation template improves our
detection significance, but only marginally (an increase of
$\sim0.3\sigma$); hence, our subsequent discussion is based on the
cross-correlation with the best-matching $\h2o$ template only, which
gives a detection significance of $4.8\sigma$ at the same $V_{\rm
  sys}$ and $K_{\rm p}$ value found with the overall best-matching
template. We note here that the data were also analysed with the SVD
method described by \citet{deK13} and the results were the same with
similar errors, suggesting that the SVD method still works well even
in more telluric-contaminated spectra than at $2.3~\mu$m. In order to
determine the line contrast ratio of the $\h2o$ lines (i.e. the depth
of the deepest $\h2o$ lines with respect to the continuum divided by
the stellar flux), we followed the method of \citet{Brog13}, injecting
an inverse scaled version of the best-matching $\h2o$ model such that
the signal at the detected planet position was exactly cancelled. This
resulted in a $\h2o$ line contrast ratio of $(1.3\pm0.2)\times10^{-3}$
for the non-convolved model. This is greater than the CO line contrast
ratio at $2.3~\mu$m ($[4.5\pm0.9]\times10^{-4}$; \citealt{deK13}) and
we found that in some cases a steep $T/P$ profile was required to
match the $\h2o$ result. This could indicate a possible overabundance
of $\h2o$ compared to CO, or possibly that the high-opacity haze
detected at optical wavelengths continues to partially obscure the CO
line depths in the $K$ band. However, within our errors we also find a
range of $T/P$ profiles where the two molecules fit to the same $T/P$
profile, meaning that with only these two CRIRES detections, we cannot
constrain the gas abundances independently from the $T/P$ profiles. In
order to constrain the $T/P$ and abundances further, a full retrieval
including both secondary eclipse and transit measurements is required,
but is beyond the scope of this Letter. Future high-resolution
observations at a wavelength where the signals of several molecules
are strong enough to be detected simultaneously will allow much
tighter constraints on the relative abundance ratios, because the
gases will be reliant on the same $T/P$ profile and continuum level,
and will likely probe overlapping regions of pressure
\citep{deK13}. Importantly, such measurements would also remove any
degeneracy with time-dependent factors, such as weather
\citep{Brog13}.

Our analysis found no increase in the cross-correlation strength when
including methane in the model spectrum, in both absorption and
emission for local thermodynamic equilibrium (LTE) chemistry. Here, we
assess claims of non-LTE methane emission at the $F_{\rm p}/F_{\rm
  s}\sim0.9$ per cent level in the atmosphere of HD 189733 b arising
from observations with SpeX on NASA's IRTF at $\sim3.25~\mu$m with an
effective resolving power of $R\sim30$ \citep{Swa10}. Such signals can
be caused by fluorescence (radiative pumping by incident photons) or
other disequilibrium processes. Support for the result was recently
published by the same group using new observations with the same
instrument at both $3.3$ and $2.3~\mu$m, and similarly strong emission
was reported in both regions at an effective resolution of $R\sim175$
\citep{Wal12}. However, \citet{Mandell11}, who observed the system
with NIRSPEC on Keck II at a resolving power of $R\sim27~000$ ruled
out emission features in the $L$ band with upper limits 40 times
smaller than expected based on the SpeX results. In a similar approach
to \citet{Mandell11}, we note that non-LTE emission lines will not be
significantly broadened by collisions in the exoplanet atmosphere, and
thus the emission intensity must be brighter at higher spectral
resolving power. At the resolving power of CRIRES ($R\sim100~000$), we
would expect to see line emission $>>1$ per cent. To test this, we
first shifted the residuals of our spectra after running {\sc Sysrem}
to the rest frame of the planet based on our detected $K_{\rm p}$,
then summed them over time (weighting each column by the number of
pixels that had not been masked in that column) to create a stacked 1D
spectrum. The largest positive deviation on detectors 1 and 3 in the
stacked spectrum (for columns where more than half of the pixels were
not masked) was $<0.6$ per cent, and the standard deviation across
both chips was $\lesssim0.1$ per cent. Hence, in agreement with
\citet{Mandell11}, our high-resolution spectra do not validate the
claims of non-LTE emission at $3.25~\mu$m.

\section*{Acknowledgements}

We would like to thank the VLT/CRIRES night astronomers and telescope
operators for their help in conducting our programme, Elena Valenti at
ESO User Support for her timely and helpful response, and our
anonymous referee for their insightful comments.

\bibliographystyle{mn2e}
\bibliography{referencesjlbh2o}{}

\label{lastpage}
\end{document}